\documentclass{ws-ijmpe}

\begin{document}

\markboth{Schunck and Dudek}{Nuclear Tetrahedal Symmetry}

%
\catchline{}{}{}{}{}
%

\title{NUCLEAR TETRAHEDRAL SYMMETRY}

\author{\footnotesize NICOLAS SCHUNCK}

\address{Department of Physics, University of Surrey, \\
Guildford, GU2 7XH, United Kingdom\\
N.Schunck@surrey.ac.uk}

\author{\footnotesize JERZY DUDEK}

\address{Group of Theoretical Physics, Institut de Recherches Subatomiques, and
Louis Pasteur University, 23 rue du Loess\\
Strasbourg 67037, France\\
Jerzy.Dudek@ires.in2p3.fr}

\maketitle

\begin{history}
\received{\today}
\revised{\today}
\end{history}

\begin{abstract}
We recall the main features of the $T^D_d$ (tetrahedral) symmetry in atomic 
nuclei and 
present realistic mean-field calculations supporting the existence such a
symmetry all over the nuclear chart. A few potential candidate-nuclei are
investigated and the possible experimental signatures of the tetrahedral
symmetry are also briefly discussed.
\end{abstract}

\section{Introduction}

It has been recently pointed out\cite{1} that atomic nuclei with  tetrahedral
symmetry could be encountered all over the nuclear chart. The predictions are
based on a very general analysis of symmetries of the nuclear mean-field and
are inspired by the group-theory considerations. The implied unique 4-fold
degeneracies of the single-particle levels characteristic of tetrahedral
(and/or octahedral) symmetry of the fermionic mean-field are thought to favour
the appearance of large gaps in the shell structure thus leading to stable
potential minima with the corresponding symmetry. [This mechanism is similar to
that of the stabilizing mechanism related to the (2j+1)-fold degeneracies at
spherical shapes]. Calculations performed for a few candidate-nuclei in various
mass regions show indeed that the tetrahedral-shape isomers are relatively low
in energy. They are separated from the ground-state minimum by a significant
barrier (up to a couple of MeV). In this paper, we would like to focus on the
nuclei whose hypothetical tetrahedral-symmetry states can be found in the
dedicated experiments; we present examples of related mean-field calculations
and briefly discuss the envisaged experimental challenges. 

\section{Theoretical Arguments Favouring the Tetrahedral Symmetry}

In the following, the nuclear shape is parametrized using the standard expansion
onto the basis of spherical harmonics $Y_{\lambda\mu}(\theta,\varphi)$:
\begin{equation}
       R(\theta, \varphi) 
       = 
       R_{0}c(\{ \alpha \} 
       \left[ 
            1 + \sum_{\lambda=2}^{\lambda_{max}}
                \alpha_{\lambda\mu} Y_{\lambda\mu}(\theta,\varphi)  
       \right]\,; \quad R_0=r_0\,A^{1/3}\,.
\end{equation}

In this expression, $R(\theta,\varphi)$ represents the nuclear surface in
spherical coordinates, $r_{0}$ is the nuclear radius parameter, $c(\{ \alpha \}
$ accounts for the volume conservation when deforming the nucleus, and
$\alpha_{\lambda\mu}$ are the deformation parameters. There exists {\it a
priori} an infinite number of ways to generate a nuclear shape with the
tetrahedral symmetry\footnote{In other words: a shape which is invariant under
every symmetry operation of the group $T_{d}^{D}$.}. Various possible
combinations of the deformation parameters as well as a detailed discussion of
the geometrical properties of the implied shapes can be found
elsewhere\cite{2}. However, the lowest multipole-order to realize the symmetry
in question is $\lambda=3$ and one can demonstrate that any other allowed
multipolarity must be greater or equal $\lambda=7$. Thus it is believed that
the high multipolarities may contribute only neglegibly. In this paper we
consider tetrahedral symmetry realized by $\alpha_{32} \neq 0$ and every other
$\alpha_{\lambda\mu}$ vanishing.

Single-particle levels show strong variation in function of $\alpha_{32}$ and
it has been demonstrated using realistic mean-field calculations that huge gaps
may appear in the corresponding spectra, comparable to, or larger than the
known spherical gaps. The predicted\cite{1,2} {\em tetrahedral-magic} numbers
for protons and  neutrons are:
\begin{center}
       N = 16, 20, 32, 40, 56, 70, 90, 112, 136$\;$\\
       Z = 16, 20, 32, 40, 56, 70, 90, 112, 126.
\end{center}
The corresponding nuclei should in principle be the best candidates for the
tetrahedral symmetry. However, pairing effects and/or quasi-particle
excitations can favour tetrahedral minima also in neighbouring nuclei. In
addition, some nuclei are likely to present other shape isomers that compete
with the tetrahedral one. It is therefore possible that the best candidates to
look for may {\em not} be those with the tetrahedral magic numbers but in their
vicinity. Case-by-case thorough calculations are therefore needed: here we
limit ourselves to presenting a few illustrations only.

Let us focus on the gaps $40$ and $70$. Two types of mean-field calculations
have been performed. One of them is based on the macroscopic-microscopic
approach where the energy is given by the sum of a liquid drop term and the 
shell- supplemented by a pairing-energy terms. In such an approach, the nuclear
shape is an input, and to find physical solutions one has to minimize the
energy as a function of the deformation parameters. Another type, the
self-consistent Hartree-Fock technique based on an effective (Skyrme) 
interaction do not explicitely predefine any dependence on the deformation, and
the deformation parameters are extracted from the selfconsistent solutions: it
is thus an output of the theory. However, the single-particle degeneracies in a
nucleus do not depend on which mean-field is used, but exclusively on the
particular symmetry of the nucleus. Indeed calculations show that the four-fold
degeneracies in the spectra are independent of the type of the mean-field used,
while the exact values of the tetrahedral gaps of course do depend somewhat on
the model and its parametrisation\footnote{The situation is very similar to
that of  spherical nuclei, where the magic numbers are the same for all kinds
of mean-fields except for the very large particle numbers (e.g. superheavy
nuclei).}.

\begin{figure}[th]
\centerline{\psfig{file=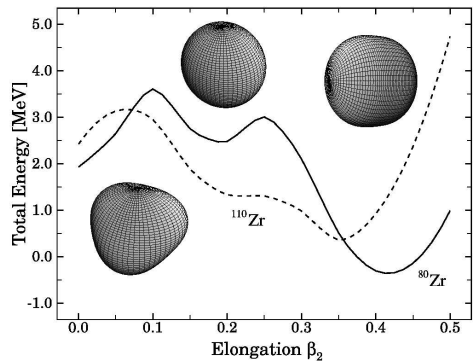,width=5cm,angle=-90}}
\vspace*{8pt}
\caption{Total energy in $^{80}$Zr and $^{110}$Zr as a function of the
         elongation of the nucleus. At each point $\beta_{2}$, the energy is 
         minimized over $\gamma$ and all octupole and hexadecapole deformation 
         degrees of freedom.}
\label{figure}
\end{figure}

In Figure \ref{figure}, we show the total energy of two isotopes of Zirconium as
a function of the elongation $\beta_{2}$. At each point, the energy is minimized
over the $\gamma$ angle and all octupole and hexadecapole degrees of freedom 
using the gradient method. The energy is calculated in the
macroscopic-microscopic approach, with a liquid drop term including curvatures
effects\cite{3}, a shell correction\cite{4} based on the non-relativistic 
Woods-Saxon potential with the universal parametrization\cite{5} and a pairing 
correction\cite{3} with particle-number projection (before variation). Two 
minima clearly appear in both nuclei: a prolate ground-state and a minimum at 
$\beta_{2} = 0$ which turns out to be tetrahedral (with $\alpha_{32} \sim 0.15$).
These calculations are very much consistent with those published
earlier\cite{6}, the latter obtained independently using self-consistent
Skyrme-Hartree-Fock technique.

We performed also Skyrme-Hartree-Fock-Bogoljubov calculations in the  region of
$^{110}$Zr, the results of which are presented in Table \ref{table}. The most
striking result shown in the Table is that with the parameterization SLy4, 
Ref.\cite{7}, the tetrahedral minima appear to be the ground-state in both 
$^{110}$Zr and $^{112}$Zr. We have verified using other Skyrme
parametrizations that the tetrahedral minima persist although they lie above the
corresponding ground-states.
\begin{table}[pt]
\tbl{HFB solutions, SLy4 parametrization, in heavy Zr isotopes for various 
     energy minima.}
{\begin{tabular}{@{}ccccc@{}} \toprule
Nucleus & Tetrahedral & Spherical & Prolate & Oblate \\\colrule
$^{108}$Zr\hphantom{00} &  0.0 & +0.391\hphantom{0}  & -1.099\hphantom{0} & -0.679\hphantom{0} \\
$^{110}$Zr\hphantom{00} &  0.0 & +0.431\hphantom{0}  & +0.072\hphantom{0} & +0.266\hphantom{0} \\
$^{112}$Zr\hphantom{00} &  0.0 & +0.027\hphantom{0}  & +0.299\hphantom{0} & +1.006\hphantom{0} \\ 
\botrule
\end{tabular}}
\label{table}
\end{table}

\section{Experimental Signatures of Tetrahedral States}

Tetrahedral minima may lead to shape isomerism; they correspond to a non-axial
octupole shape. Consequently, if rotational bands are built on top of these
configurations they should lead to parity doublets. However, contrary to the
better-known case of axial-octupole deformations, no E1 inter-band transitions
should be observed due to the vanishing dipole moment. The bands should be
sought at relatively low spins since the angular momentum alignment will tend
to break the tetrahedral symmetry, and at relatively high excitation energy.
Indeed, tetrahedral minima are predicted at about 1-2~MeV above the
ground-states in most of the cases; the corresponding moments of inertia  are
expected to be smaller than those in the prolate minima, the intraband decay
energies should be correspondingly larger. Also: the single-particle energies
in a tetrahedral nucleus beeing up to 4-fold degenerate, one may expect to
observe a unique 16-fold, approximately degenerate, quasi-particle pattern.

We have presented results of microscopic calculations supporting the 
existence of very low-lying tetrahedral shape minima in Zirconium isotopes. 
The eventuality of tetrahedral ground-states in the heaviest Zr must also be 
considered. These nuclei are not beyond the range of the current experimental 
facilities.

\end{document}